\documentclass[runningheads]{llncs}
\usepackage[T1]{fontenc}

\usepackage{amsmath}
\usepackage{enumitem}
\usepackage{cleveref}
\usepackage[most]{tcolorbox} 
\usepackage{graphicx} 

\usepackage{float}
\usepackage{listings}
\usepackage{xcolor}

\definecolor{codegreen}{rgb}{0,0.6,0}
\definecolor{codegray}{rgb}{0.5,0.5,0.5}
\definecolor{codepurple}{rgb}{0.58,0,0.82}
\definecolor{backcolour}{rgb}{0.95,0.95,0.92}

\lstdefinestyle{mystyle}{
    backgroundcolor=\color{backcolour},   
    commentstyle=\color{codegreen},
    keywordstyle=\color{magenta},
    numberstyle=\tiny\color{codegray},
    stringstyle=\color{codepurple},
    basicstyle=\ttfamily\footnotesize,
    breakatwhitespace=false,         
    breaklines=true,                 
    captionpos=b,                    
    keepspaces=true,                 
    numbers=left,                    
    numbersep=5pt,                  
    showspaces=false,                
    showstringspaces=false,
    showtabs=false,                  
    tabsize=2
}

\newfloat{code}{htbp}{loc}
\floatname{code}{Listing}

\crefname{code}{Listing}{Listings}
\Crefname{code}{Listing}{Listings}   

\usepackage{fancyvrb}
\newenvironment{cverb}
  {\VerbatimEnvironment
   \begin{center}
   \begin{BVerbatim}}
  {\end{BVerbatim}
   \end{center}}

\begin{document}
\title{BOOP: Write \textit{Right} Code}
%
%
\author{Vaani Goenka \and
Aalok Thakkar}
\authorrunning{V. Goenka and A. Thakkar}
%
\institute{Ashoka University}
\maketitle              
\begin{abstract}


Novice programmers frequently adopt a syntax-specific and test-case-driven approach, writing code first and adjusting until programs compile and test cases pass, rather than developing correct solutions through systematic reasoning. AI coding tools exacerbate this challenge by providing syntactically correct but conceptually flawed solutions.
In this paper, we address the question of developing correctness-first methodologies to enhance computational thinking in introductory programming education. To this end, we introduce BOOP (Blueprint, Operations, OCaml, Proof), a structured framework requiring four mandatory phases: formal specification, language-agnostic algorithm development, implementation, and correctness proof. This shifts focus from making the code {\em work} to understanding {\em why} the code is correct.

BOOP was implemented at Ashoka University using a VS Code extension and preprocessor that enforces constraints. Initial evaluation shows improved algorithmic reasoning and reduced trial-and-error debugging. Students reported better edge case understanding and problem decomposition, though some initially found the format verbose. Instructors observed stronger foundational skills compared to traditional approaches, suggesting that structured correctness-first approaches may significantly improve students' computational thinking abilities.
\end{abstract}

\section{Introduction}
\label{sec:intro}

Computer Science, at its core, is a discipline of abstraction, logic, and problem-solving. Writing code is a powerful tool to express these ideas. Yet, we frequently note that students beginning their journey in computer science at Ashoka University conflate the tool with the discipline and adopt a predominantly program-centric mindset. This manifests in several ways: students incorrectly confuse the expected behaviour of the program on a small set of test cases with total functional correctness, attempt to mechanically patch incorrect programs instead of reasoning from first principles, and develop a syntax-specific mindset in programming. In general, students even tend to overestimate their understanding of program constructions~\cite{ref_article2}. This is reflected in their prioritisation of {\em coding a solution} over understanding the underlying algorithm. Roundabout approaches often result in students only being able to consider test cases that are over-fitted to their code, disregarding the problem's specifications and overall correctness. The ACM 2023 guidelines for undergraduate computer science education explicitly recommend instruction in proof-writing and correctness verification, including loop invariant application for algorithm correctness and inductive methods for recursive algorithm analysis~\cite{ref_article21}. 

This is more prominent in the age of AI-based programming tools and co-pilots. Students unfamiliar with the computational thinking process are tempted to consult language models for answers to programming problems~\cite{ref_article32}. However, language models churn out poor-quality code, and sometimes even inaccurate statements, which leads to poor programming practices~\cite{ref_article14,ref_article37}. This worsens the student’s skill in and outlook on programming. They develop misconceptions about the necessity of learning problem-solving skills when tools such as  Large Language Models (LLMs) could generate solutions for them, trusting these tools much more than they call for~\cite{ref_article14}. 

It is imperative to shift the focus in introductory CS courses from `writing a program' to `writing {\em right} programs.' This paper explores how instructors can adopt a novel computer science pedagogy paradigm that promotes accurate best-practice programming by design.
\section{Related Work}
\label{sec:related-work}

Computing education research has extensively documented the challenges students face when learning introductory programming. Students step in with preconceptions about programming difficulty or develop negative attitudes early in their learning journey, leading to ineffective learning approaches such as rote memorization and, in some cases, academic dishonesty.

A common pedagogical intervention is reducing extraneous cognitive load while preserving the essential problem-solving aspects of programming instruction. Constrained environments like mini-languages (turtle graphics) and block-based systems (MIT's LogoBlocks~\cite{ref_article3}) mimic basic programming concepts like achieving a goal by breaking it down into a correct sequence of instructions. These approaches use syntactic constraints as scaffolding and support ``hybrid programming environments that blend block-based and text-based tools'' \cite{ref_article4}. 

However, fundamental programming concepts must not be over-abstracted to the point  that application and formalization become disconnected. While referential frameworks can provide an initial conceptual understanding, students' learning and evaluation should be guided by a more robust and well-structured system that comprises both formal and informal methods.

\subsection{Functional Programming Education}

Research identifies the most challenging aspects of programming for students as: (1) debugging programs, (2) designing programs to solve specific problems, and (3) decomposing functionality into procedures, all of which are fundamental to functional programming~\cite{ref_article2,ref_article36}. Additionally, students reported that 2 of their top 3 difficulties were error handling and recursion. Moreover, instructors highlighted the need for more problem-solving-oriented pedagogical materials~\cite{ref_article2,ref_article10}. On the topic of problem-solving, Hudak, a pioneer of the International Haskell Committee, emphasizes the importance of developing problem specifications before implementation details while acknowledging the need for effective compilation techniques to make functional languages practical for education~\cite{ref_book1}. These are key ideas that guide our research. 

\subsection{AI and CS Pedagogy}

Recent research has examined the intersection of artificial intelligence tools and programming education, revealing significant pedagogical challenges. While LLMs perform well with high-resource languages like Python, their capabilities with low-resource languages such as OCaml remain limited~\cite{ref_article13}. Even when ChatGPT provides correct solutions to functional programming problems in OCaml 68\% of the time, only half of these solutions are pedagogically valuable or ``benefit students in some form''\cite{ref_article14}. Moreover, ChatGPT-generated code exhibits detectable patterns that even remain consistent across prompt variations\cite{ref_article13}. Accordingly, researchers recommend emphasizing code comprehension, debugging, and correctness within structured design frameworks~\cite{ref_article32,ref_article33}.

\subsection{Pattern-Based Approaches to Programming Education}

Educational research has also explored pattern-based pedagogical frameworks such as POSA and GoF~\cite{ref_book2,ref_article16}. These patterns encapsulate successful solution strategies by describing problem contexts, competing forces, and balancing structures. Although these ideas are valuable, they take away from the autonomy and worthwhile struggle of a student. A pedagogical approach must emphasize independent pattern discovery and understanding over pattern memorization, prioritizing the development of transferable algorithmic thinking skills.

\subsection{Correctness-by-Construction}

Several frameworks for stable, correct-by-construction programming have emerged. Meyer's Design By Contract is a systematic approach to designing programs where preconditions, postconditions, and invariants are explicitly defined for each component~\cite{ref_article29}. Wing explicitly mentions that computational thinking involves using ``invariants to describe a system's behaviour succinctly and declaratively''~\cite{ref_article30}. Gries and Schneider emphasize the use of a rigorous, equational style of proving and show how logic can be applied to formally develop programs from specifications~\cite{ref_article31,ref_book3}. Loksa et al. investigated meta-cognitive awareness in novice programmers by identifying six problem-solving stages and explicitly coaching students to recognize which stage they occupied when encountering difficulties~\cite{ref_article35}. A study which leveraged their framework to investigate common errors found that successful students differed primarily in forming correct initial problem conceptualization, corresponding to the first stage: \textit{reinterpret the problem prompt} \cite{ref_article34}. There remains a need to integrate formal specification methods with novice-friendly structured learning environments for CS education.
\newtcolorbox{boxedquote}{
  colback=white,        
  colframe=black,       
  boxrule=1pt,          
  arc=0pt,              
  left=6pt, right=6pt,  
  top=6pt, bottom=6pt   
}

\section{Overview}
\label{sec:overview}

In this section, we demonstrate our framework through a representative programming task from the Introduction to Computer Science course at our institution. The selected exercise exemplifies the challenges inherent in introducing formal specification-driven programming to CS students:

\begin{boxedquote}
The natural numbers can be defined as: \texttt{type nat = Zero | Succ of nat}. Define a function \verb|div| such that it takes two natural numbers \verb|a| and \verb|b|, and returns the quotient and the remainder when \verb|a| is divided by \verb|b|.
\end{boxedquote}

This problem operates within a constructive type system based on Peano arithmetic, where natural numbers are represented inductively. The type \verb|nat| compels students to explicitly engage with recursive decomposition and constructive reasoning that is often implicit in conventional arithmetic operations. The given problem statement is deliberately made to omit several details: it provides no guidance on handling edge cases such as division by zero (whether through error handling, option types, or alternative approaches), does not specify the explicit type signature for the \verb|div| function, and does not offer concrete test cases. This ambiguity forces students to engage in specification refinement as a prerequisite to implementation, requiring them to formalise their understanding of the problem domain before attempting a solution.

\subsection{Three Incorrect Approaches}

We first analyse student submissions for this question. Three representative submissions are presented in Appendix \ref{app:incorrect_approaches}. We observe the following misconceptions: 

\begin{enumerate}
\item \textbf{Imperative State Tracking:} Students rightly but inadequately conceptualize division as iterative subtraction, tracking only the quotient, failing to account for the remainder.
\item \textbf{Incomplete Edge Case Handling:} Implementations exhibit unhandled division by zero due to conflation of validity of statements with their mathematical contexts.
\item \textbf{Lack of Constructive Foundation:} Students incorrectly invoke \verb|int| operators for \verb|nat| types, demonstrating incomplete understanding of custom types, type safety, and program requirements.
\end{enumerate}

\subsection{Pedagogical Analysis}

\begin{code}[t]
\begin{lstlisting}[language=caml]
type nat = Zero | Succ of nat

let plus (a: nat) (b: nat) : nat =
  match a with
  | Zero -> b
  | Succ a' -> Succ (plus a' b)

let mult (a: nat) (b: nat) : nat = 
  match b with 
  | Zero -> Zero 
  | Succ b' -> plus (mult a b') a

let less_than (a: nat) (b: nat) : bool = 
  match (a, b) with 
  | (_, Zero) -> false 
  | (Zero, Succ _) -> true
  | (Succ a', Succ b') -> less_than a' b'
\end{lstlisting}
\caption{Definition of addition, multiplication, and comparison on custom-defined natural numbers in OCaml.}
\label{lst:overview-plus-mult}
\end{code}

These errors reveal critical junctures in the learning process. The intuitive notion of addition as repeated successors or multiplication as repeated addition translates seamlessly into programming (as shown in \cref{lst:overview-plus-mult}). That is, students can conceptualise \verb|mult| as iteratively adding \verb|a| to itself \verb|b| times, and the implementation follows naturally from this mental model. 

Division, however, is not merely repeated subtraction. It requires simultaneously tracking how many times the subtraction occurred (the quotient). This dual-tracking requirement fundamentally challenges students' sequential thinking patterns. Moreover, students frequently carry over mathematical intuitions that do not translate directly into program logic. The type \verb|nat| is bounded below by \verb|Zero|, requiring all arithmetic operations to explicitly handle edge cases. Unlike multiplication (which is safe within the \verb|nat| domain), subtraction and division require explicit safety checks and totality considerations. Furthermore, students tend to underspecify edge cases: while division by zero is rarely encountered in ordinary mathematical contexts, programming demands exhaustive case analysis. This transition from partial mathematical reasoning to total computational specification constitutes a fundamental epistemological shift.

\subsection{Blueprint Specification}

Addressing this shift requires explicit mathematical specification as a pedagogical intervention. Leslie Lamport compares the lack of specifications to building ``a skyscraper without a blueprint''~\cite{ref_article22}. 
And yet, programmers routinely construct complex systems without formal specifications~\cite{ref_article22,ref_article23,ref_article24,ref_article25}. Research in CS education consistently shows that students who engage with formal preconditions and postconditions develop robust problem-solving skills~\cite{ref_article29,ref_article30,ref_article31}. 

By requiring students to articulate precise mathematical relationships before implementation, we address what Soloway and Spohrer (1989) identify as the {\em specification problem}: the tendency for novice programmers to conflate problem understanding with solution construction~\cite{ref_article26}.

For our running example, we specify the {\em blueprint}, 
that is, the correctness criteria, as a function contract. 
Function contracts, rooted in Hoare's axiomatic semantics~\cite{ref_article27}, express computational behaviour through requires clauses or preconditions (what must be true before execution) and ensures clauses or postconditions (what will be true after execution), forming a logical triple $\{P\} S \{Q\}$ where $P$ is the precondition, $S$ is the statement, and $Q$ is the postcondition. This formalism compels students to reason explicitly about program correctness before engaging with the implementation details.
\\\\
To write a correct \verb|div| function, we specify the following:

\begin{enumerate}
    \item \textbf{Requires:} \verb|a : nat|, \verb|b : nat|, and \verb|b <> Zero| (that is \verb|b| is non-zero).
    \item \textbf{Ensures:} Returns a pair \verb|(q, r)| such that:
    \begin{enumerate}
        \item \verb|a = plus (mult b q) r|
        \label{ensures:div-inv}
        \item \verb|less_than r b = true|
        \label{ensures:div-lt}
    \end{enumerate}
\end{enumerate}

For clarity and convenience, we will overload familiar arithmetic operators such as \verb|+|, \verb|*|, and \verb|<| with the \verb|nat| operations as presented in \cref{lst:overview-plus-mult}.

\subsection{Algorithm as Operations}
\label{sec:overview:operations}

The algorithmic insight follows Gries' methodology in {\em The Science of Programming}~\cite{ref_book3}: treat the first ensures clause \verb|a = (b * q) + r| as an invariant that must be maintained throughout the computation, while systematically working toward satisfying the second clause \verb|r < b| through repeated subtraction. This transforms the abstract division specification into concrete operation steps:

\begin{enumerate}
    \item With \verb|q = Zero| and \verb|r = a|, the first ensures clause (\ref{ensures:div-inv}) holds. In case if \verb|a < b|, then both the ensures clauses hold and the function can return \verb|q = Zero| and \verb|r = a|.
    \item If \verb|a < b|, then since division is fundamentally repeated subtraction, we {\em subtract} \verb|b| from \verb|a| to get \verb|a'| such that \verb|a' + b = a|. We need to define a subtraction operator that can perform this step.
    \item Now suppose our function \verb|div| can correctly divide \verb|a'| by \verb|b| and obtain a quotient \verb|q'| and remainder \verb|r'| such that 
    \begin{enumerate}
        \item \verb|a' = (b * q') + r'|
        \item \verb|r' < b|.
    \end{enumerate}
    Assuming the correctness of \verb|plus| and \verb|mult| operators, we construct the solution to the original problem: since \verb|a = a' + b| and \verb|a' = (b * q') + r'|, we have:
    \begin{cverb}
a = ((b * q') + r') + b
  = (b * q') + (r' + b) 
  = (b * (Succ q')) + r'
    \end{cverb}
    
    Therefore, the quotient for \verb|div a b| is \verb|Succ q'| and the remainder is \verb|r'|. 
\end{enumerate}

\begin{code}[t]
\begin{lstlisting}[language=caml]
let rec minus (n: nat) (m: nat) : nat =
  match n, m with 
  | _, Zero -> n
  | Zero, _ -> failwith "minus: function requires n >= m"
  | Succ n', Succ m' -> minus n' m'

let rec div (a: nat) (b: nat) : (nat * nat) =
  match b with
  | Zero -> failwith "div: division by zero"
  | _ -> safe_div a b

and safe_div (a: nat) (b: nat) : (nat * nat) =
  match (less_than a b) with
  | true -> (Zero, a)
  | false -> 
      let (q', r') = safe_div (minus a b) b in (Succ q', r')
\end{lstlisting}  
\caption{A translation of the operational steps for division and safe subtraction on natural numbers in OCaml syntax.}
\label{lst:overview-div}
\end{code}

These operation steps capture the crux of algorithmic thinking in computer science: the translation of abstract specifications into precise, language-agnostic procedures. They demonstrate how to {\em systematically derive a correct implementation from mathematical reasoning}. The operational steps can be translated faithfully into any programming paradigm, whether functional, imperative, or object-oriented, because they encode the fundamental logical structure of the division algorithm. In \cref{lst:overview-div} we present it in OCaml.

This solution handles all cases appropriately, maintains readability, and lends itself to formal verification. One can easily adapt these operations for imperative implementations using loops and mutable variables if functional programming is not the target paradigm. Moreover, this algorithmic structure generalizes beyond natural numbers to any Euclidean domain where division with remainder is well-defined. The core operations of comparison, subtraction, and iterative reduction remain conceptually identical.

We note that such an approach may not always be the most efficient in terms of computational complexity. In our pedagogical framework, we deliberately prioritize correctness over efficiency. This emphasis aligns with established principles in CS education: Knuth argued that premature optimisation is the root of all evil~\cite{ref_article39} and emphasized that programs should be written primarily for human comprehension, with efficiency as a secondary consideration~\cite{ref_article1}. Research by Soloway et al. demonstrates that novice programmers who focus on correctness first develop stronger debugging skills and produce robust and more maintainable code~\cite{ref_article26}. 

\subsection{Proof of Correctness}

We have designed the code to be correct by construction, following the mathematical specification directly through recursive decomposition. While testing can be used to identify errors, and indeed, {\em to show the presence of bugs}, it can never demonstrate their absence. 
In contrast, a formal proof establishes that the specification holds for all valid inputs, regardless of how many test cases are considered. In this case, the correctness of the function \verb|div| is not merely a design intuition or observed behaviour but a theorem to be proven:

\begin{theorem}
The function \verb|div| in \cref{lst:overview-div} correct; that is, for given natural numbers \verb|a : nat| and \verb|b : nat|:
\begin{enumerate}
    \item If \verb|b = Zero|, then the function raises a runtime error (division by zero).
    \item If \verb|b <> Zero|, then \verb|div a b| returns a pair \verb|(q, r)| such that:
    \begin{enumerate}
        \item \verb|a = (b * q) + r|, and
        \item \verb|r < b|.
    \end{enumerate}
\end{enumerate}
\end{theorem}

This theorem asserts that the program satisfies its specification (blueprint) in all well-defined cases. To establish this, we proceed by structural induction, analysing the recursive structure of the function and verifying that both postconditions are preserved at each step. A complete proof, as expected of students in our Introduction to Computer Science course, is provided in Appendix~\ref{app:proof}.

This methodology exemplifies our approach to teaching systematic program development. We want students to learn this disciplined thinking process:
\begin{enumerate}
    \item Establish a precise {\bf blueprint} through a function contract that has a requires and an ensures clause (or a loop contract with invariants).
    \item Develop language-agnostic {\bf operational steps} that describe the algorithm.
    \item Translate these steps into program syntax (in our case, in {\bf OCaml}) while preserving the logical structure.
    \item Provide a {\bf proof} that guarantees correctness.
\end{enumerate}

This sequence embodies the principle of correct-by-construction programming, where careful specification and methodical decomposition lead naturally to verifiable implementations. By internalizing this systematic approach, students develop the discipline necessary for tackling complex computational problems with confidence and rigour, moving beyond ad-hoc programming, towards principled software development practices. 
\section{The BOOP Framework for Systematic Problem-Solving}
\label{sec:boop}

The methodology demonstrated in the previous section forms the foundation of our broader framework: BOOP (Blueprint, Operations, OCaml, Proof). This structured approach draws inspiration from Knuth's concept of Literate Programming, which emphasized that programs should be written as literature intended for human understanding rather than mere machine instruction. As Knuth articulated: ``Instead of imagining that our main task is to instruct a computer what to do, let us concentrate rather on explaining to human beings what we want a computer to do.'' \cite{ref_article1}.

Literate programming emphasizes structured design with clear separation of concerns, human-first ordering that prioritizes cognitive comprehension over execution sequence, and meta-language abstraction that builds logical constructs before implementation details. 
This shift toward human-centric program development aligns with our goals of fostering language-agnostic algorithmic thinking. 

\subsection{The Four-Phase BOOP Framework}

The BOOP framework operationalizes these principles through a mandatory four-phase structure for student submissions:

\begin{enumerate}
    \item \textbf{Blueprint:} Students establish specifications, that is, a functional correctness criterion (preconditions and postconditions). The next iteration of our implementation will support the option for students to add representative test cases as well as state asymptotic complexity bounds. 
    \item \textbf{Operations:} They then design detailed algorithmic steps, expressed in natural language with minimal mathematical notation.
    \item \textbf{OCaml Code:} Students finally translate their operational steps into an implementation, adhering to constructive programming principles and explicit type annotations. Currently, we support OCaml.
    \item \textbf{Proof:} Students finally provide a rigorous proof through structural induction or invariant analysis, demonstrating that their implementation satisfies the blueprint specification. This phase closes the loop between specification and verification, embodying the correct-by-construction methodology.
\end{enumerate}

This four-phase BOOP framework prevents the premature leap from problem statement to implementation that characterizes novice programming behaviour. By mandating explicit articulation at each phase, students develop reflective computational thinking practices. The framework's language-agnostic emphasis prepares students for professional software development's polyglot reality, where algorithmic insights must transcend programming language syntax. 

The BOOP acronym, where the second O stands for `OCaml code' was chosen because the course at our institution is taught in OCaml, and because of the acronym's appeal and memorability among students. Currently, BOOP supports OCaml but it is a language-agnostic framework for problem-solving and can be adapted to support other languages.

\subsection{Enforcement Through Technical Infrastructure}

\subsubsection{Program Preprocessor}
Based on experience from Spring 2025, students have been increasingly submitting AI-generated code. This circumvents learning objectives while appearing syntactically correct. The experience motivated us to develop a preprocessor that promotes conceptual understanding through instructive feedback. Our tool parses OCaml submissions before compilation, identifying counterproductive patterns (such as the use of built-in functions like \verb|List.rev| instead of native implementations, imperative constructs like mutation and variable reassignment that contradict functional programming paradigms, nested functions defined in local scope that obscure algorithmic structure, and functions with anonymous inputs) that prevent clear reasoning about program behaviour.

Furthermore, our system restricts specific syntax and practices that undermine learning goals: it encourages recursive thinking, mandates explicit type reasoning through typed function definitions, promotes first-principles implementation over excessive built-in function usage, and requires pattern matching where it would be more appropriate than if-then-else control flow. These constraints are configurable and can be dynamically adjusted based on assignment objectives, student proficiency levels, and classroom instruction progress. However, while the framework allows for substantial customisation by the instructor, it is not a requirement. Assignments can be deployed with liberal default settings. 

\subsubsection{Integrated Development Environment}

On the technical side, the only requirement is installing VS Code, which we found was already the preferred IDE for most students. A custom VS Code extension integrates the BOOP framework with an instructor interface for configuring assignments, specialized syntax highlighting, a completion check to enforce all four phases, and evaluation tools for targeted, conceptual feedback. 

The integrated environment transforms the traditional code-compile-test cycle into a specification-design-implement-verify workflow. We create a learning ecosystem where good practices become the path of least resistance, fostering sustainable computational thinking skills.

\section{Observations and Discussion}

\subsection{Student Learning Outcomes} 

In the ongoing iteration of the Introduction to Computer Science course, 21 students were asked to submit their solutions to an assignment consisting of functional programming questions using BOOP. The assignment submissions in the previous iteration of the course comprised only of OCaml code and sometimes typeset proofs written in \LaTeX. We acknowledge our sample size is too small to draw conclusive trends.  

Our preliminary evaluation of BOOP reveals significant pedagogical benefits in structured programming assignments. A majority of students demonstrated rigorous algorithmic reasoning and proof skills, with participants successfully constructing sound and complete proofs using the framework's components. This contrasts sharply with previous cohorts who struggled with proof structure and identification of correctness criteria. 

Students reported that BOOP's scaffolding addressed critical gaps in programming instruction. One participant noted that conventional online tutorials ``just start writing code like everyone knows what is to be done,'' whereas BOOP provided explicit problem decomposition strategies absent from other learning resources. There were also students who struggled with one part but showed a good grasp of the others. For example, one student could not write code properly but demonstrated good conceptual understanding through blueprint and operational step components. This separation enabled targeted feedback and revealed that coding difficulties often masked sound algorithmic reasoning. 

Even incomplete solutions demonstrated adherence to sound programming principles, with students avoiding discouraged practices such as inappropriate built-in function usage. This suggests that BOOP's preprocessor and emphasis on specification promote disciplined implementation approaches, ultimately encouraging a sustainable thought process over obtaining solutions. Students particularly valued the framework's ability to clarify the transition between pseudocode and implementation. The structured separation of base and recursive cases addressed a common source of confusion in functional programming. As one student observed, ``without BOOP I would forget about correctness and not really think of edge-cases.'' 

These findings suggest that explicit structural frameworks can significantly improve novice programmers' formal reasoning abilities while maintaining code quality standards, despite require rigorous elaboration. A student said ``BOOP takes four times the time, but I understand the problem much better now.'' The framework's emphasis on correctness-first thinking may help address persistent challenges in computing education related to formal methods integration and long-term computational thinking development.

\subsection{Implementation Challenges and Student Misconceptions}

Despite these positive outcomes, our evaluation revealed several pedagogical challenges that warrant attention in future implementations. 

A significant challenge emerged in students' approach to the Blueprint component. Rather than developing the correctness required by the problem statement, some students still engaged in code-first approaches, subsequently reverse-engineering the blueprint. Students produced \texttt{ensures} clauses that merely restated the implementation logic (sometimes even using the exact if-else clauses written in their functions) rather than abstracting the function's intended behaviour or establishing correctness criteria independent of their implementation. A student said, ``Blueprint just requires you to rewrite the question'', and could not appreciate the skill of developing formal correctness statements.

Our observations show the importance of explicit instruction in formal methods concepts and their practical relevance. It also suggests the need for making lectures tailored to the same approach that assignment submissions entail.

As a part of the evaluation, we also consulted CS educators with decades of experience in teaching introductory courses. All of the consulted educators recognised value in our framework as it encouraged rigorous articulation of requirements and set the appropriate tone and momentum for a computer science trajectory. Instructors in advanced programming courses offered in our institution (and otherwise~\cite{ref_article38}) encounter students with inadequate programming practices and weak methods. Early intervention through approaches like BOOP may address these persistent educational challenges and make students well-prepared for future coursework. 
We continue to explore multi-institutional deployments, comparative analysis with traditional programming pedagogies, longer-term retention studies, and industry preparation assessment. 
\section{Conclusion and Future Directions}

Our work demonstrates that BOOP offers a structured, first-principles approach to computer science education, emphasising reasoning and correctness, over syntax-specific and test-case driven approaches. Several promising research directions emerged from our preliminary evaluation:

\begin{enumerate}
    \item Integrate BOOP with automated theorem proving and verification environments capable of generating counterexamples and evaluating student-constructed proofs. Tools such as Dafny, Lean, or Why3 are scalable and provide immediate feedback on proof correctness while automatically generating edge cases that challenge student solutions, to establish a comprehensive pipeline supporting automated assessment and systematic feedback. As an easier alternative, be combined with online judges for test-case verification, coming closer to current traditional modes of checking the accuracy of code against a problem statement.
    \item To facilitate framework adoption after an initial phase of resistance among students, we plan to tailor lectures to mirror the problem-solving paradigm of BOOP, in an attempt to discourage the notion that BOOP is a final formatting step for assignment submissions only. The perception of BOOP as verbose often stems from students who reverse-engineer correctness criteria, leading to repetition rather than genuine problem analysis. To address this, lectures must emphasise that BOOP components are problem-solving tools, not mere submission requirements. Since students have favored instant feedback, a potential direction is also to use non-suggestive, LLM-driven prompts. We can rate-limit these to discourage frivolous use.
\end{enumerate}

As AI reshapes the landscape of software development, the programmers who will thrive are not those who can compete with machines in code generation, but those who can harness computational thinking to solve problems that machines cannot even formulate. The BOOP structure prepares students for this future by teaching them to think like computer scientists first, and programmers second.

\subsubsection*{Acknowledgements.} We gratefully acknowledge the Mphasis AI \& Applied Tech Lab at Ashoka University for supporting this research.

\appendix

\section{Three Incorrect Approaches for the Division Function}
\label{app:incorrect_approaches}

\subsection{Imperative State Tracking:}

\begin{lstlisting}[language=caml]
let rec minus (n : nat) (m : nat) : nat = 
  match (n, m) with 
  | (Zero, _) -> Zero
  | (_, Zero) -> n 
  | (Succ n', Succ m') -> minus n' m'

let rec div1 (a: nat) (b: nat) : nat = 
  match a with 
  | Zero -> Zero 
  | _ -> Succ (div1 (subtract a b) b)
\end{lstlisting}

This approach reflects the students' tendency to conceptualise division as iterative subtraction, attempting to track a single accumulating value (quotient) rather than the requisite quotient-remainder pair. The three key errors in the function \verb|div1| are: 
\begin{enumerate}
    \item Incorrect quotient if there is a remainder (three divided by two returns two)
    \item Division by zero is not handled, and leads to unbounded recursion
    \item The remainder is not tracked. 
\end{enumerate}

These are caused due to a malformed recursive structure that fails to correctly decompose the problem into its constituent mathematical components.

\subsection{Incomplete Edge Case Handling:}

\begin{lstlisting}[language=caml]
let rec div2 (a: nat) (b: nat) : (nat * nat) = 
  match a with
  | Zero -> (Zero, Zero)
  | Succ a' -> 
      let (q, r) = div2 a' b in
      if (Succ r) < b then (q, Succ r)
      else (Succ q, Zero)
\end{lstlisting}

The \verb|div2| function demonstrates a partial understanding of the problem structure, but exhibits critical flaws in specification completeness. Students correctly identify the need for a pair return type but fail to address division by zero, assume the availability of comparison operators without formal definition, and conflate structural recursion on the dividend with the mathematical properties of division.

\subsection{Lack of Constructive Foundation:}

\begin{lstlisting}[language=caml]
let div3 (a: nat) (b: nat) : (nat * nat) = (a / b, a mod b)
\end{lstlisting}

The function \verb|div3| reveals the difficulty in transitioning from abstract mathematical operations to constructive algorithmic implementations. Students invoke \verb|int| operators for \verb|nat|, demonstrating an incomplete understanding of custom types, type safety, and the nature of the task.

\clearpage

\section{Proof of Correctness for the Division Function}
\label{app:proof}

\begin{theorem}
The function \verb|div| in \cref{lst:overview-div} correct; that is, for given natural numbers \verb|a : nat| and \verb|b : nat|:
\begin{enumerate}
    \item If \verb|b = Zero|, then the function raises a runtime error (division by zero).
    \item If \verb|b <> Zero|, then \verb|div a b| returns a pair \verb|(q, r)| such that:
    \begin{enumerate}
        \item \verb|a = (b * q) + r|, and
        \item \verb|r < b|.
    \end{enumerate}
\end{enumerate}
\end{theorem}

\begin{proof} 
We prove the correctness of the function \verb|div| by considering cases:

\begin{enumerate}
    \item \textbf{Case 1:} \verb|b = Zero|. According to the definition of \verb|div| in \cref{lst:overview-div}, line 9:
    \begin{cverb}
match b with Zero -> failwith "div: division by zero"
    \end{cverb}
    the function immediately raises a runtime error.

    \item \textbf{Case 2:} \verb|b <> Zero|. In this case, \verb|div a b| calls \verb|safe_div a b|. We proceed by strong induction on \verb|a : nat|, that is, we assume the correctness of the function for all strictly smaller inputs \verb|a' < a| to prove it for \verb|a|.

    \begin{description}
        \item[\textbf{Base case:}] Let \verb|a = Zero|.
        Then \verb|Zero < b| since \verb|b <> Zero|, and thus:
        \begin{cverb}
safe_div Zero b -> (Zero, Zero)            
        \end{cverb}
        by pattern matching on \verb|less_than a b|.  Hence, \verb|div Zero b| returns \verb|(Zero, Zero)|. We verify the two postconditions:
        \begin{enumerate}[label=(\alph*)]
            \item \verb|(b * Zero) + Zero = Zero|, so \verb|a = (b * q) + r| holds.
            \item \verb|Zero < b|, as required.
        \end{enumerate}

        \item[\textbf{Inductive Hypothesis:}] For all \verb|a' : nat| such that \verb|a' < a|, assume that \verb|safe_div a' b| returns \verb|(q', r')| satisfying:
        \begin{enumerate}[label=(\alph*)]
            \item \verb|a' = (b * q') + r'|, and
            \item \verb|r' < b|.
        \end{enumerate}

        \item[\textbf{Inductive Step:}] Consider the evaluation of \verb|safe_div a b|.

        \begin{enumerate}[label=(\alph*)]
            \item \textbf{If} \verb|a < b|: 
            Then the function returns \verb|(Zero, a)|. We check the postconditions:
            \begin{enumerate}[label=(\roman*)]
                \item \verb|(b * Zero) + a = a|, so where \verb|q = Zero|, \verb|r = a|.
                \item \verb|a < b|, by assumption.
            \end{enumerate}

            \item \textbf{If} \verb|a < b = false|: 
            Then the function evaluates:
            \begin{cverb}
let (q', r') = safe_div (minus a b) b in (Succ q', r')
            \end{cverb}

            Note that \verb|(a - b)| is well-defined since \verb|a < b = false|. Also, we have \verb|(a - b) < a| for all \verb|b <> Zero|, so by the inductive hypothesis applied to \verb|a' = a - b|:
            \[
            \verb|a - b = plus (mult b q') r'|, \quad \text{and} \quad \verb|r' < b|.
            \]

            Adding \verb|b| to both sides yields:
            \[
            \verb|a = (a - b) +  b = ((mult b q') +  r') +  b|.
            \]

            By correctness of addition and multiplication operations: 
            \[
            \verb|a = (b * (Succ q')) + r'|.
            \]

            Therefore, the returned pair \verb|(Succ q', r')| satisfies:
            \begin{enumerate}[label=(\roman*)]
                \item \verb|a = (b * (Succ q')) + r'|,
                \item \verb|r' < b|.
            \end{enumerate}
        \end{enumerate}
    \end{description}
\end{enumerate}
\qed
\end{proof}

\clearpage

\end{document}